\documentclass[12pt]{article}
\input{epsf}

\usepackage{graphicx}
\font\bm=cmmib10 at 10pt
\font\bms=cmmib10 at 7pt \textfont9=\bm \scriptfont9=\bms
\usepackage{amsfonts}
\mathchardef\balpha= "790B
\mathchardef\bbeta= "790C
\mathchardef\bTheta= "7902
\mathchardef\bzeta= "7910
\mathchardef\bOmega= "790A
\mathchardef\bGamma= "7900
\mathchardef\bDelta= "7901
\mathchardef\bPhi= "7908
\mathchardef\bphi= "791E
\mathchardef\bomega= "7921
\mathchardef\bxi= "7918
\mathchardef\bet= "7911
\mathchardef\brho= "791A
\mathchardef\btau= "791C
\mathchardef\bmu= "7916
\mathchardef\bvarpi= "7924
\mathchardef\btheta= "7912

\def\perthou{^o\!\!/_{\!\!oo}}

\def \lvec{(\kern-.26em(}
\usepackage{amsmath}
\usepackage{amssymb}
\usepackage{authblk}
\usepackage{url}
\def\pmb#1{\setbox0=\hbox{#1}%
\def \lvec{(\kern-.26em(}
\kern-.025em\copy0\kern-\wd0
\kern.05em\copy0\kern-\wd0
\kern-.025em\raise.0433em\box0 }
\begin{document}
\title{Variations of atmospheric $^{12}$CO$_2$ and $^{13}$CO$_2$  from the year 1000 to 2024}
\author[1]{ W. A. van Wijngaarden}
\author[2] {W. Happer}
\affil[1]{Department of Physics and Astronomy, York University, Canada}
\affil[2]{Department of Physics, Princeton University, USA}
\renewcommand\Affilfont{\itshape\small}
\date{\today}
\maketitle
\begin{abstract}
A simple model gives good agreement between observed emissions of CO$_2$ into the atmosphere from the combustion of fossil fuels and observed increases of the atmospheric concentrations $C$ of CO$_2$ and decreases of the isotope delta value $\delta^{13}{\rm C}$.  If fossil fuel emissions were to cease the model predicts that: (1) the difference between the  CO$_2$ concentration $C$ and the preindustrial value, $C_0\approx 280$ ppm, would decay exponentially with time constant $\tau_1\approx 60$ y; (2) the difference between the concentration-delta-value product $C\times\delta^{13}{\rm C}$ and the preindustrial value  $C_0\times\delta^{13}{\rm C}_0
=(280 \hbox{ ppm})\times(-6.5\,\perthou)$, would decay exponentially with a time constant $\tau_2\approx 8 $ y.  The decay times are different because of isotopic exchange, largely driven by photosynthesis and respiration on the land.
\end{abstract}
%\keywords{Suess effect, fossil fuel emissions, isotope exchange}
\section{Introduction}
As is widely known, and  summarized in a recent paper by Engelbeen, Hannon and Burton\,\cite{EHB},
the increasing concentration $C$ of atmospheric CO$_2$ and the decrease of the isotope delta value $\delta^{13}$C (the Suess effect), can be qualitatively explained by the addition of $^{13}$C-depleted CO$_2$ from the combustion of fossil fuels to the atmosphere. In this note, we show that both facts are quantitatively consistent with a simple, 2-parameter model. 
If there were no further emissions of fossil fuels, the first parameter is the exponential decay time $\tau_1$ of any difference of the CO$_2$ concentration $C$ from  the preindustrial value, $C_0 =280$ ppm. The second parameter is the exponential decay time $\tau_2$ of any difference of the concentration-delta-value product $C\times\delta^{13}{\rm C}$ from the preindustrial value, $C_0\times\delta^{13}{\rm C}_0 =280\hbox{ ppm}\times  (-6.5\,\perthou)$.
\begin{figure}[h]
\begin{centering}
%\postscriptscale{emis.eps}{1}
\includegraphics[height=80mm,width=.8 \columnwidth]{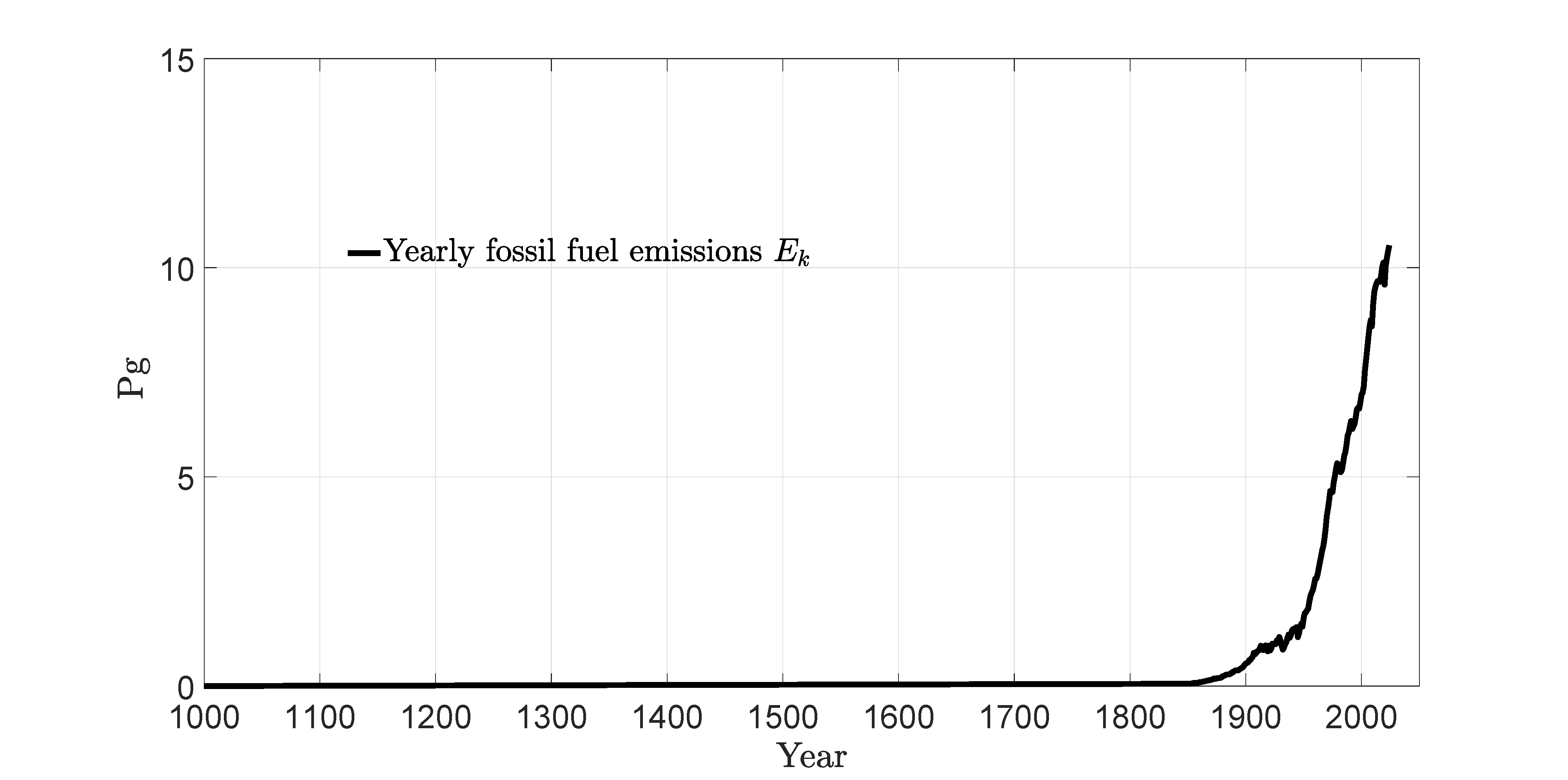}
\caption{Yearly fossil fuel emissions of CO$_2$ to the atmosphere in petagrams (Pg) or $10^{15}$ g calendar year. The observational data is from reference\,\cite{emissions,Fried}. }
\label{emis}
\end{centering}
\end{figure}
\begin{figure}[h]
\begin{centering}
%\postscriptscale{Cdel.eps}{1}
\includegraphics[height=80mm,width=.8\columnwidth]{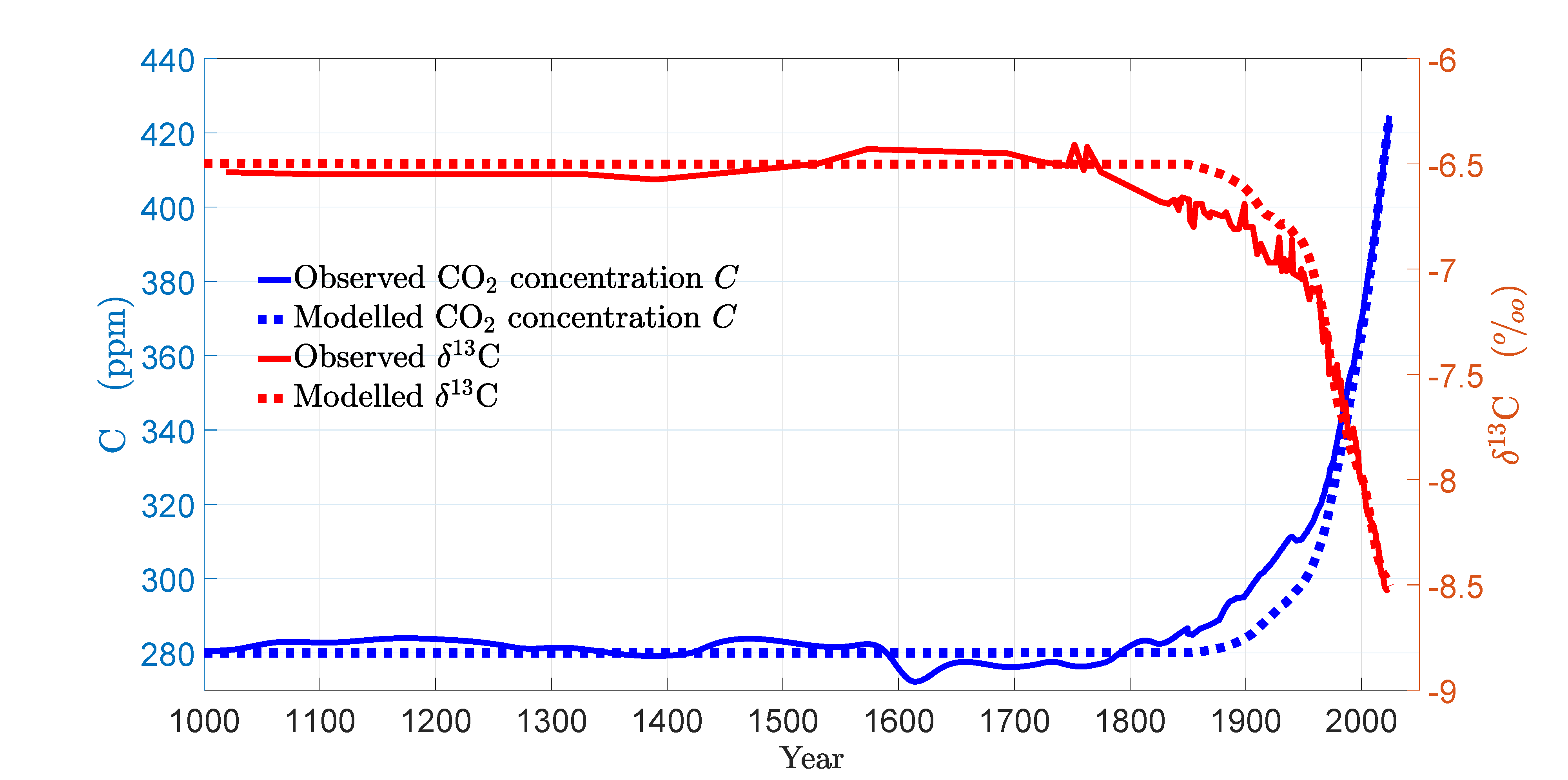}
\caption{The continuous blue line shows observed yearly averages of CO$_2$ concentrations $C=N/N_{\rm at}$ in ppm (parts per million by volume) obtained from references \cite{Lan, Etheridge,Keeling0,MacFarling}. The continuous red line is observed isotope delta values $\delta^{13}$C found from references \cite{Rubino1,Rubino2,Keeling}. The corresponding dotted lines are
model predictions from (\ref{d12}) and (\ref{d14}). The preindustrial concentrations of CO$_2$  were close to  $C_0= 280$ ppm and the preindustrial isotope delta values were close to $\delta^{13}{\rm C}_0 = -6.5\,\perthou$. The e-folding time constant $\tau_1=1/\gamma_1$ of  (\ref{d12}) for the decay of concentration perturbations, $C-C_0$, was $\tau_1= 60$ y.  The e-folding time constant $\tau_2=1/\gamma_2$ of  (\ref{d14}) for the decay of perturbations, $C\times \delta^{13}{\rm C}-C_0\times\delta^{13}{\rm C}_0$, of concentration-delta-value products was  $\tau_2= 8$ y. } 
\label{Cdel}
\end{centering}
\end{figure}

\section{A Simple Model}

\subsection{Total carbon}
Let the atmosphere contain  $N_1$  atoms of the stable carbon isotope $^{12}$C and   $N_2$  atoms of the stable isotope $^{13}$C. Most of these atoms will be in 
CO$_2$ molecules.  To facilitate subsequent discussions, we will describe the numbers of carbon atoms as the elements of a $(2\times 1)$ dimensional {\it isotope number vector} 
\begin{equation}
|N\rangle = \left [\begin{array}{l}N_1\\ N_2\end{array}\right]=N\left [\begin{array}{l}n_1\\ n_2\end{array}\right].\label {m4}
\end{equation}
Here
\begin{equation}
 N=N_1+N_2,\quad n_1 = N_1/N,\quad n_2=N_2/N\quad \hbox{and}\quad n_1+n_2 = 1.\label {m6}
\end{equation}
Large scale additions of  CO$_2$ to the atmosphere began around the year 1850, as shown in Fig. \ref{emis}.
From Fig. \ref{Cdel} we see that the total number $N$ of  carbon atoms, or the concentration $C=N/N_{\rm at}$, was nearly constant before the year 1800. The total number of atmospheric molecules\,\cite{Alexander} is
\begin{equation}	
N_{\rm at}=1.065\times 10^{44}.
\label{m16}
\end{equation}
The pre-industrial concentration of CO$_2$ was about
\begin{equation}
C_0 =  280 \hbox{ ppm}.  \label {m14}
\end{equation}
We denote the preindustrial numbers $P_1$ of $^{12}$C atoms and $P_2$ of $^{13}$C atoms with an isotope number  vector analogous to (\ref{m4}),
\begin{equation}
|P\rangle = \left [\begin{array}{l}P_1\\ P_2\end{array}\right]=P \left [\begin{array}{l}p_1\\ p_2\end{array}\right],\label {m8}
\end{equation}
where
\begin{equation}
 P=P_1+P_2,\quad p_1 = P_1/P,\quad p_2=P_2/P\quad \hbox{and}\quad p_1+p_2 = 1.\label {m10}
\end{equation}
The total number of preindustrial carbon atoms in the atmosphere was
\begin{equation}
P = P_1+P_2=N_{\rm at} C_0= 2.98 \times 10^{40}.  \label {m12}
\end{equation}
We used the preindustrial CO$_2$ concentration $C_0$  from (\ref{m14}) and the total number of atmospheric molecules $N_{\rm at}$ of (\ref{m16}) to write the numerical value on the right of (\ref{m12})

 Let $F_1$ denote the addition rate of  $^{12}$C atoms from the combustion of fossil fuels, and let $F_2$ the addition rate of $^{13}$C atoms.  In analogy to (\ref{m4}) and (\ref{m8}), we will write these as the fossil-fuel {\it isotope source vector}
\begin{equation}
|F\rangle = \left [\begin{array}{l}F_1\\ F_2\end{array}\right]=F\left [\begin{array}{l}f_1\\ f_2\end{array}\right],\label {m17a}
\end{equation}
where
\begin{equation}
 F=F_1+F_2,\quad f_1 = F_1/F,\quad f_2=F_2/F\quad \hbox{and}\quad f_1+f_2 = 1.\label {m17b}
\end{equation}
The elements of $|F\rangle$ have the units of atoms per unit time. For the elements of $|N\rangle$ and $|P\rangle$ the units are simply atoms.

Reference \cite{HE} showed that the observed number, $N = N(t)$,  of atmospheric CO$_2$ molecules at time $t$ is  given accurately in terms of emissions $F'=F(t')$ at earlier times $t'$ by the simple mathematical expression
\begin{equation}
N=\int_{-\infty}^t dt' e^{-\gamma_1 (t-t')} F' +P. \label {m20}
\end{equation}
Differentiating (\ref{m20}) with respect to time, we see that it is a solution of the differential equation
\begin{eqnarray}
\frac{d}{dt} N&=&-\gamma_1\int_{-\infty}^t dt'  e^{-\gamma_1 (t-t')}F'+ F\nonumber\\
&=&-\gamma_1(N-P)+F .\label {m24}
\end{eqnarray}
For no fossil fuel emissions, $F=0$, the differential equation (\ref{m24}) is an analog of {\it Newton's law of cooling}\,\cite{Newton}.  If the number $N$ of CO$_2$ differs from the preindustrial value $P$, and if there are no additions $F$ of CO$_2$ from fossil fuel combustion (if $F=0$), then $N$ will decay exponentially at the rate $\gamma_1$ to the preindustrial value $P$.

For time-independent conditions, when $dN/dt = 0$ and $dF/dt =0$ or $F = $ constant, the solution to (\ref{m24}) is
\begin{equation}
N=\frac{P\gamma_1 +F}{\gamma_1}=\frac{D+F}{\gamma_1}. \label {m26}
\end{equation}
Here we have introduced a {\it deep source} of CO$_2$ that is injected  into the atmosphere,
\begin{equation}
D=P\gamma_1 = 4.97\times 10^{38}\hbox{ y}^{-1}. \label {m28}
\end{equation}
The numerical value on the right of (\ref{m28}) came from assuming the decay rate
\begin{equation}
\gamma_1=1/60\hbox{ y}. \label {m29}
\end{equation}
The numerical value (\ref{m29}) of $\gamma_1$ was used to draw the model curves of Fig. \ref{Cdel}, as we will describe below.
The mean mass of a terrestrial carbon atom is
\begin{equation}
m=1.994\times 10^{-23} \hbox{ g}. \label {m30}
\end{equation}
Using the numerical value (\ref{m30}) with (\ref{m28}), we find that the pre-industrial addition rate of carbon to the atmosphere, in mass per unit time, is 
\begin{equation}
\frac{dM}{dt} = Dm  = 9.91 \hbox{ Pg y}^{-1}. \label {m32}
\end{equation}
Here Pg is the unit petagram, or $10^{15}$ g of carbon atoms.  From inspection of the top panel of Fig. \ref{emis}, we see that the current emission rate of fossil fuel carbon, around 10.5 Pg y$^{-1}$ in the  year 2024, is slightly larger than the model preindustrial rate (\ref{m32}).

To facilitate  subsequent analysis, we introduce the row vector 
\begin{equation}
\langle U_1|=\big[1\quad 1\big] . \label {m34}
\end{equation}
We note that $N=N_1+N_2$, the sum of the number $N_1$ of $^{12}$C  atoms
and  $N_2$ of $^{13}$C atoms, can be written as the product of the row vector (\ref{m34}) and the column vector (\ref{m4}),
\begin{eqnarray}
\langle U_1|N\rangle = \begin{array}{r r}[1& 1]\\ & \end{array}\left [\begin{array}{l}N_1\\ N_2\end{array}\right]=N_1+N_2 = N.
\label {m36}
\end{eqnarray}
In like manner we can use (\ref{m34}) with (\ref{m8}) and (\ref{m17a}) to write
\begin{equation}
 P=\langle U_1|P\rangle\quad\hbox{and}\quad  F=\langle U_1|F\rangle.  \label {m38}
\end{equation}
Letting
\begin{equation}
\gamma = \gamma_1, \label {m40}
\end{equation}
we use (\ref{m36}) and (\ref{m38}) to write the last line of (\ref{m24}) as
\begin{eqnarray}
\frac{d}{dt}\langle U_1|N\rangle =-\gamma_1\langle U_1| \bigg (|N\rangle -|P\rangle\bigg)+\langle U_1|F\rangle .\label {m42}
\end{eqnarray}

\subsection{Isotope fractions}
The stable isotope $^{12}$C  makes up about 98.9\% of Earth's carbon. The heavier isotope, $^{13}$C, makes up almost all of the remaining 1.1\%. 
Although the ratio  $N_2/N_1$ of the number $N_2$ of $^{13}$C atoms  to the number $N_1$ of $^{12}$C atoms is very nearly $0.011$,  {\it isotopic fractionation}\,\cite{IF} can cause slight differerences in the isotope ratios of different carbon samples.  For example, the organic carbon in trees has a lesser fraction of  $^{13}$C  than the air from which the carbon was extracted and fixed by photosynthesis. This is because the lighter $^{12}$CO$_2$ molecules diffuse slightly faster through leaf stomata and through the leaf tissue to the chloroplasts. And the carbon-fixing enzyme rubisco works slightly more efficiently with the lighter carbon isotope.

It is customary to quantify the ratio, $ [N_2/N_1]_x$, in a sample $X$ of carbon atoms by  the {\it isotope delta value}
\begin{equation}	
\delta^{13}{\rm C}=\delta_{xv}=\frac{[N_2/N_1]_x}{[N_2/N_1]_v}-1.
\label{if10}
\end{equation}
Mass spectrometers are used to precisely measure the ratio
$[N_2/N_1]_x$ in the  sample $X$, with respect to $[N_2/N_1]_v$ in a calibrating standard $V$.  The standard is usually {\it Vienna Pee Dee Belemnite} or VPDB. This standard traces back to fossil belemnite shells from a late Cretaceous formation along the Pee Dee river in South Carolina\,\cite{PD}. Contemporary VPDB standards are artificially prepared to have isotopic ratios nearly identical to the original belemnite standards.  A recent measurement\,\cite{aPD} found that the absolute ratio of $^{13}$C to $^{12}$C atoms in VPDB standards is

\begin{equation}	
[N_2/N_1]_v = 0.0111105 \pm 0.0000042.
\label{if11}
\end{equation}
The pre-industrial value of $\delta^{13}{\rm C}$ for the atmosphere, determined from the CO$_2$ of air trapped in ice core bubbles, is around
\begin{equation}	
\delta _{pv}= -6.5\,\perthou.
\label{if12}
\end{equation}
where the symbol $\perthou$ denotes parts per thousand, or the number $10^{-3}$.
 The value of $\delta^{13}{\rm C}$ for CO$_2$ from fossil fuel emissions averages about
\begin{equation}	
\delta_{fv}= -28\,\perthou.
\label{if14}
\end{equation}
For the year 2026, the observed  $\delta^{13}{\rm C}$ for the atmosphere was 
\begin{equation}	
\delta_{nv}= -8.7\,\perthou.
\label{if14a}
\end{equation}
The fact that (\ref{if14a}), the contemporary value of $\delta^{13}{\rm C}$, is less than the preindustrial value (\ref{if12}) is what one would qualitatively expect from the addition $^{13}$C-depleted carbon from fossil fuels to the atmosphere. This is called the Suess effect\,\cite{Suess}. More discussion of the isotope delta value  $\delta^{13}{\rm C}$ can be found in the Appendix.

For further analysis, it will be convenient to introduce a row vector $\langle U_2|$, analogous to $\langle U_1|$ of (\ref{m34}), and defined in terms of the fractions $p_1$ of $^{12}$C and $p_2$ of $^{13}$C in the preindustrial atmosphere by
\begin{equation}
\langle U_2|=\big[-p_2\quad p_1\big] . \label {if16}
\end{equation}
Then we see that

\begin{eqnarray}	
\langle U_2|N\rangle &=& \begin{array}{r r}[-p_2& p_1]\\ & \end{array} \left [\begin{array}{l}N_1\\ N_2\end{array}\right]\nonumber\\
& =& -p_2N_1+p_1N_2\nonumber\\
&=& p_2N_1\left(-1+\frac{p_1N_2}{p_2N_1}\right)\nonumber\\
&=& Np_2n_1\delta_{np}\nonumber\\
&=& Np_1p_2\delta_{np}+\mathcal{O}(\delta^2).
\label{if20}
\end{eqnarray}
To write the last line of (\ref{if20}) we noted from the Appendix that the fraction $n_1$ of $^{12}$C atoms in the current atmosphere is related to the fraction $p_1$ in the preindustrial atmosphere by $n_1=p_1+\mathcal{O}(\delta)$. Therefore, to order $\delta$ we can use (\ref{if20}) to write
\begin{equation}	
\delta_{np}= \frac{\langle U_2|N\rangle}{p_1p_2 N}.
\label{if22}
\end{equation}
From inspection of Fig. \ref{Cdel} one can see that   the isotope delta value,  $\delta^{13}$C, remained close to the preindustrial value $\delta^{13}{\rm }C = -6.5\,\perthou$ from the year 1000 to the year 1800, the beginning of the industrial age, when fossil fuel combustion began to add significant amounts of $^{13}$C-depleted CO$_2$ to the atmosphere.  It is therefore natural to assume that the variable $\langle U_2|N\rangle$ of (\ref{if22}), which is proportional to $N\delta_{np}$, obeys an analog of Newton's law of cooling\,\cite{Newton}, or an analog of (\ref{m42}) for the number $N$ of CO$_2$ molecules in the atmosphere,
\begin{equation}
\frac{d}{dt}\langle U_2|N\rangle =-\gamma_2\langle U_2|\bigg(|N\rangle -|P\rangle\bigg)+\langle U_2|F\rangle .\label {if24}
\end{equation}
\subsection{Reciprocal vectors}
To facilitate further discussion we define column  vectors,
\begin{equation}
|U_1\rangle = \left [\begin{array}{l}p_1\\ p_2\end{array}\right]\quad\hbox{and}\quad|U_2\rangle = \left [\begin{array}{r}-1\\ 1\end{array}\right], \label {rv4}
\end{equation}
that are the reciprocals of the row vectors $\langle U_1|$ of (\ref{m34}) and $\langle U_2|$ of (\ref{if16}),
\begin{equation}
\langle U_1|=\big[1\quad 1\big] \quad\hbox{and}\quad \langle U_2|=\big[-p_2\quad p_1\big]. \label {rv2}
\end{equation}
The reciprocal row and column vectors are readily seen to be orthonormal,
\begin{equation}
\langle U_j|U_k\rangle =\delta_{jk}, \label {rv6}
\end{equation}
where the Kronecker delta symbol is $\delta_{jk}=1$ if $j=k$ and $\delta_{jk}=0$ if $j\ne k$. 
The vectors are also complete in the sense that
\begin{equation}
\sum_j |U_j\rangle\langle U_j| = \hat 1=\left [\begin{array}{cc}1&0\\ 0&1\end{array}\right]. \label {rv7}
\end{equation}
Here we have used the symbol $\hat 1$ to denote the $(2\times 2)$ identity matrix. One can verify that the product of the identity matrix $\hat 1$ with an arbitrary column vector $|X\rangle$ on the right or an arbitrary row vector $\langle X|$ on the left simply reproduces the vectors
\begin{equation}
\hat 1|X\rangle=|X\rangle\quad \hbox{and}\quad \langle X|\hat 1 = \langle X|. \label {rv7a}
\end{equation}
To prove the completeness (\ref{rv7}) of the reciprocal vectors we use (\ref{rv4}) and (\ref{rv2}) to write
\begin{eqnarray}
\sum_j |U_j\rangle\langle U_j|  &=& \left [\begin{array}{l}p_1\\ p_2\end{array}\right]\!\!\begin{array}{r r}[1& 1]\\ & \end{array}
 + \left [\begin{array}{r}-1\\ 1\end{array}\right]\!\!\begin{array}{r r}[-p_2& p_1]\\ & \end{array}\nonumber\\
&=& \left [\begin{array}{ll}p_1&p_1\\ p_2&p_2\end{array}\right]
 + \left [\begin{array}{rr}p_2&-p_1\\ -p_2&p_1\end{array}\right]\nonumber\\
&=& \left [\begin{array}{ll}1&0\\0&1\end{array}\right]=\hat 1.
\label {rv8}
\end{eqnarray}
To write the last line of (\ref{rv8}) we recalled from (\ref{m10}) that $p_1+p_2 = 1$.

Multiplying both sides of (\ref{m42}) on the left by $|U_1\rangle$, multiplying both sides of  (\ref{if24}) on the left by $|U_2\rangle$, and adding the resulting equations we find
\begin{eqnarray}
&&|U_1\rangle\frac{d}{dt}\langle U_1|N\rangle +|U_2\rangle\frac{d}{dt}\langle U_2|N\rangle =\nonumber\\
&-&|U_1\rangle\gamma_1\langle U_1\bigg(|N\rangle-|P\rangle\bigg)-|U_2\rangle\gamma_2\langle U_2\bigg(|N\rangle-|P\rangle\bigg)
+|U_1\rangle\langle U_1|F\rangle +|U_2\rangle\langle U_2|F\rangle.\label {rv10}
\end{eqnarray}
Using the completeness property (\ref{rv7}) we can write (\ref{rv10}) as 
\begin{equation}
\frac{d}{dt}|N\rangle =-\Gamma\bigg(|N\rangle-|P\rangle\bigg)+|F\rangle.\label {rv14}
\end{equation}
In (\ref{rv14}) the damping matrix $\Gamma$ is
\begin{eqnarray}
\Gamma &=&|U_1\rangle\gamma_1\langle U_1|+|U_2\rangle\gamma_2\langle U_2|\nonumber\\
& =& \left [\begin{array}{ccc}p_1\gamma_1+p_2\gamma_2&\quad&p_1(\gamma_1-\gamma_2)\\ p_2(\gamma_1-\gamma_2)&\quad& p_2\gamma_1+p_1\gamma_2\end{array}\right]\nonumber\\
& =& \left [\begin{array}{rcr}\gamma_1 +\gamma_{12} p_2&\quad&-\gamma_{21} p_1
\\ -\gamma_{12} p_2&\quad&\gamma_1+ \gamma_{21} p_1\end{array}\right].\label {rv16}
\end{eqnarray}
Here we have introduced the characteristic {\it isotope exchange rate}
\begin{equation}
\gamma_{12}=\gamma_{21} =|\gamma_2-\gamma_1|.\label {rv17}
\end{equation}
From inspection of the first line of (\ref{rv16}) and the orthonormality relation  (\ref{rv6}) we see that
the row vectors $\langle U_q|$ of (\ref{rv2}) are left eigenvectors of $\Gamma$  with eigenvalues $\gamma_q$, and the column  vectors $| U_q\rangle $ of (\ref{rv4}) are right eigenvectors   with the same eigenvalues $\gamma_q$,
\begin{equation}
\langle U_q|\Gamma =\gamma_q\langle U_q| \quad\hbox{and}\quad \Gamma|U_q\rangle = |U_q\rangle\gamma_q.\label {rv18}
\end{equation}
From inspection of (\ref{m8}) and (\ref{rv4}) we see that the preindustrial isotope number vector $|P\rangle$ 
is proportional to the right eigenvector $|U_1\rangle$ of $\Gamma$ 
\begin{equation}
|P\rangle = P|U_1\rangle.\label {rv19a}
\end{equation}
Using (\ref{rv19a}) and (\ref{rv18}), we can write (\ref{rv14}) as
\begin{equation}
\frac{d}{dt}|N\rangle =-\Gamma|N\rangle+\gamma_1|P\rangle+|F\rangle.\label {rv19b}
\end{equation}
A general solution of (\ref{rv19b}) is
\begin{equation}
|N\rangle =\int_{-\infty}^{t} dt' e^{-\Gamma (t-t')}|F'\rangle+ |P\rangle+e^{-\Gamma t}|K\rangle. \label {rv20}
\end{equation}
Here the time-independent, $(2\times 1)$ vector $|K\rangle$ must be determined by boundary conditions.
For $|K\rangle\ne 0$ the solution (\ref{rv20}) is unphysical, since it goes to infinity as $t\to -\infty$,.
We therefore set $|K\rangle=0$ and we write (\ref{rv20}), in analogy to (\ref{m20}), as 
\begin{equation}
|N\rangle =\int_{-\infty}^{t} dt' e^{-\Gamma (t-t')}|F'\rangle +|P\rangle. \label {rv20c}
\end{equation}

We recall that $e^x$, the exponentiated value of any $(2\times 2)$ matrix $x$ of  the form  
\begin{equation}
x=|U_1\rangle x_1\langle U_1|+|U_2\rangle x_2\langle U_2| \label {rv21a}
\end{equation}
can be evaluated with the Taylor series, 
\begin{eqnarray}
e^x &=&\hat 1 +\frac{x}{1!}+\frac{x^2}{2!}+\cdots\nonumber\\
&=&|U_1\rangle\langle U_1|+|U_2\rangle\langle U_2|+\frac{1}{1!}\bigg(|U_1\rangle x_1\langle U_1|+|U_2\rangle x_2\langle U_2|\bigg)
+\frac{1}{2!}\bigg(|U_1\rangle x_1^2\langle U_1|+|U_2\rangle x_2^2\langle U_2|\bigg)+\cdots\nonumber\\
&=&|U_1\rangle\bigg(1+\frac{x_1}{1!}+\frac{x_1^2}{2!}+\cdots\bigg)\langle U_1|+|U_2\rangle\bigg(1+\frac{x_2}{1!}
+\frac{x_2^2}{2!}+\cdots\bigg)\langle U_2|
\nonumber\\
&=&|U_1\rangle e^{x_1}\langle U_1|+|U_2\rangle e^{x_2}\langle U_2|.
\nonumber\\
\label {rv21b}
\end{eqnarray}
Evaluating  (\ref{rv21b}) with  $x=-\Gamma(t-t')$,  $x_1 = -\gamma_1(t-t')$ and  $x_2 = -\gamma_2(t-t')$, in accordance with (\ref{rv16}),  we see that the exponentiated matrix of (\ref{rv20}) can be writtten as
\begin{equation}
e^{-\Gamma (t-t')} =|U_1\rangle e^{-\gamma_1(t-t')}\langle U_1|+|U_2\rangle e^{-\gamma_2(t-t')}\langle U_2|. \label {rv22}
\end{equation}
Eq. (\ref{rv22}) implies that
\begin{equation}
\langle U_q|e^{-\Gamma (t-t')} = e^{-\gamma_q(t-t')}\langle U_q|\quad\hbox{and}\quad e^{-\Gamma (t-t')}|U_q\rangle =e^{-\gamma_q (t-t')}|U_q\rangle. \label {rv22a}
\end{equation}

In component form, we can use (\ref{rv16}) to write (\ref{rv19b}) as
\begin{eqnarray}
 \frac{d}{dt}\left [\begin{array}{l}N_1\\N_2 \end{array}\right] =  
 \left [\begin{array}{l}-\gamma_1N_1-\gamma_{12} p_2 N_1+\gamma_{21} p_1 N_2+\gamma_1P_1+F_1
\\ -\gamma_1N_2+\gamma_{12} p_2N_1-\gamma_{21} p_1N_2+\gamma_1P_2+F_2\end{array}\right].\label {rv24}
\end{eqnarray}
\begin{figure}[h]
%\postscriptscale{Variations.eps}{1}
\begin{centering}
\includegraphics[height=80mm,width=.8\columnwidth]{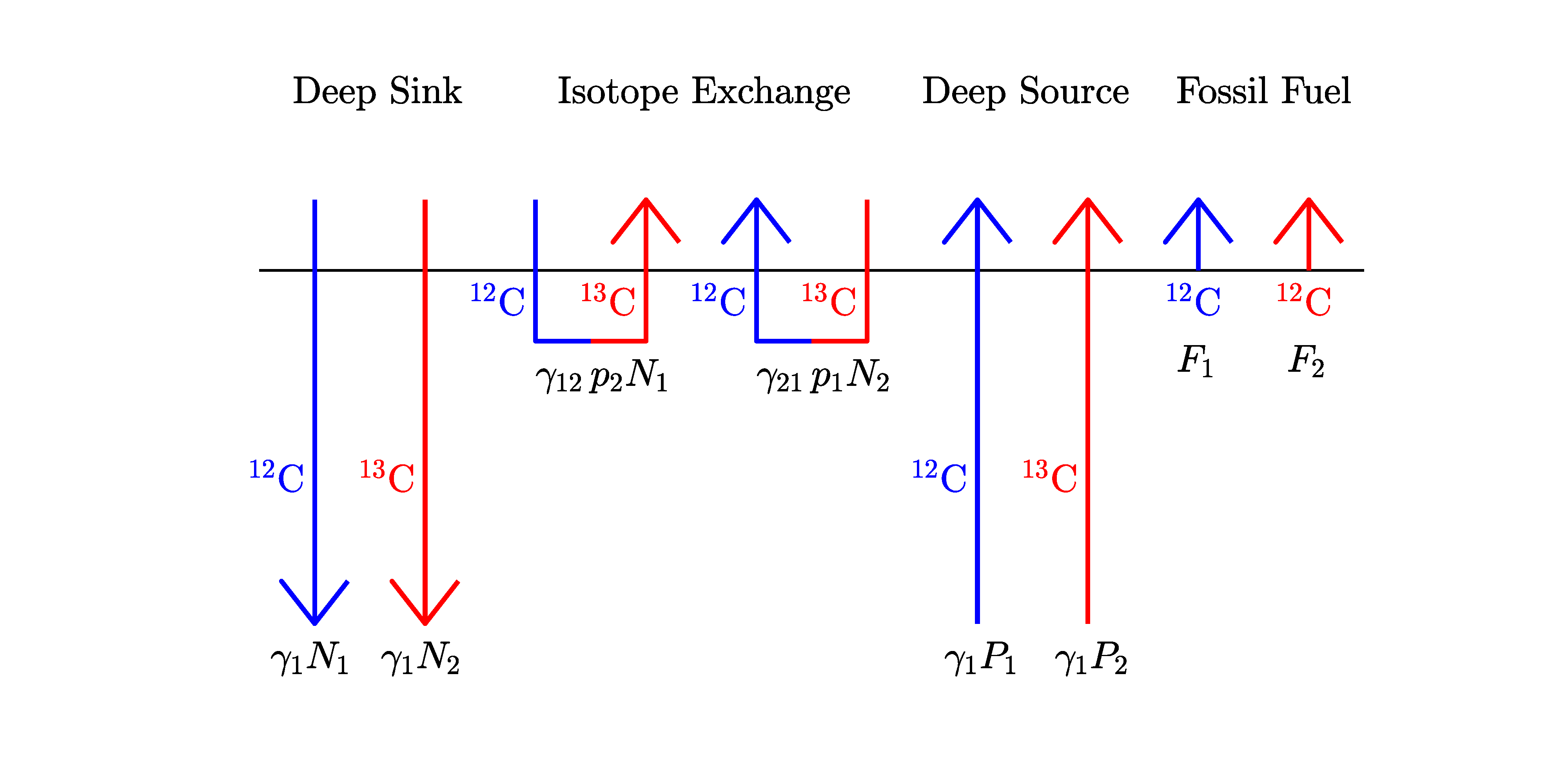}
\caption { The rates of change of carbon isotopes in the atmosphere. The most abundant isotope, $^{12}$C is shown in blue and the rare stable isotope $^{13}$C is shown in red.  Negative terms on the right of (\ref{rv24}), or {\it sinks},  represent losses of carbon atoms from the atmosphere, and are denoted by downward arrows. Positive terms on the right of (\ref{rv24}),  {\it sources}, represent carbon atoms added to the atmosphere and are denoted with upward arrows. }
\label{Variations}
\end{centering}
\end{figure}
The physical significance of the terms on the right of (\ref{rv24}) is illustrated by Fig. \ref{Variations}. Carbon atoms are transfered from the atmosphere to the deep ocean and deep terrestrial soil at the same rate, $\gamma_1$, for both isotopes, $^{12}$C and $^{13}$C.  The  isotopes are exchanged at rates proportional to the characteristic exchange rate $\gamma_{12}$ of (\ref{rv17}), and this drives isotope delta values $\delta^{13}$C toward preindustrial values. The rapid drawdown of atmospheric CO$_2$ by photosynthesis of land plants during the summer, and the steady return of carbon due to the respiration of soil and water microorganisms throughout the year makes a major contribution to the exchange rates.  Upwelling deep ocean water and oxidation of deeply buried reduced carbon in the soil both inject CO$_2$ with preindustrial isotope delta values $\delta^{13}$C into the atmosphere. 
\subsection{The Suess effect}
Multiplying both sides of (\ref{rv20}) on the left by $\langle U_1|$, and using (\ref{m36}) and (\ref{m38}) we find
\begin{eqnarray}
\langle U_1|N\rangle &=&\int_{-\infty}^t dt' \langle U_1|e^{-\Gamma (t-t')}|F'\rangle +\langle U_1|P\rangle\nonumber\\
&=&\int_{-\infty}^t dt'e^{-\gamma_1 (t-t')} \langle U_1|F'\rangle +P\nonumber\\
&=&\int_{-\infty}^t dt'e^{-\gamma_1 (t-t')} F' +P
\label {su2}
\end{eqnarray}
or 
\begin{eqnarray}
N=\Delta N_1 +P. \label {su4}
\end{eqnarray}
We used (\ref{rv22a}) to write the second line of (\ref{su2}).
The {\it fossil fuel increments} $\Delta N_q=\Delta N_q(t)$, with $q=1$ or $q=2$, are defined by
\begin{equation}
\Delta N_q=\int_{-\infty}^t dt'e^{-\gamma_q (t-t')}F'.
\label {su6}
\end{equation}
Multiplying both sides of (\ref{rv20}) on the left by $\langle U_2|$ we find,
\begin{eqnarray}
\langle U_2|N\rangle &=&\int_{-\infty}^t dt' \langle U_2|e^{-\Gamma (t-t')}|F'\rangle +\langle U_2|P\rangle\nonumber\\
&=&\int_{-\infty}^t dt'e^{-\gamma_2 (t-t')} \langle U_2|F'\rangle +\langle U_2|P\rangle.
\label {su8}
\end{eqnarray}
Noting that $\langle U_2|P\rangle= P(-p_2p_1+p_1p_2)= 0$, we use (\ref{if20}) to write (\ref{su8}),  to order $\delta$, as
\begin{eqnarray}
Np_1p_2\delta_{np}
&=&\int_{-\infty}^t dt'e^{-\gamma_2 (t-t')}F'p_1p_2\delta_{fp}.
\label {su10}
\end{eqnarray}
Dividing both sides of (\ref{su10}) by $Np_1p_2$ and using (\ref{su6}) we find
\begin{equation}
\delta_{np}=\frac{\Delta N_2}{N}\delta_{fp}.
\label {su12}
\end{equation}
As shown by (\ref{a18}), the isotope delta values are nearly additive so we can use (\ref{su12}) to write
\begin{eqnarray}
\delta_{nv}&=&\delta_{np}+\delta_{pv}\nonumber\\
&=&\frac{\Delta N_2}{N}\delta_{fp}+\delta_{pv}.
\label {su14}
\end{eqnarray}
\subsubsection{No exchange}
From inspection of (\ref{rv17}) we see that in the limit of no isotopic exchange, $ \gamma_{12}\to 0$
\begin{equation}
\gamma_2\to \gamma_1.
\label {ne2}
\end{equation}
Then we see from (\ref{su6}) and (\ref{su4})  that 
\begin{equation}
\Delta N_2\to \Delta N_1 = N-P.
\label {ne4}
\end{equation}
Substituting (\ref{ne4}) into (\ref{su14}), we find
\begin{eqnarray}
\delta_{nv}&=&\frac{(N-P)\delta_{fp}}{N}+\delta_{pv}\nonumber\\
&=&\frac{(N-P)(\delta_{fv}-\delta_{pv})+N\delta_{pv}}{N}\nonumber\\
&=&\frac{(N-P)\delta_{fv}+P\delta_{pv}}{N}.
\label {ne6}
\end{eqnarray}
To write the second line of (\ref{ne6}) we recalled from (\ref{a18}) that the shift parameters $\delta$ are nearly additive, so 
\begin{equation}
\delta_{fv} = \delta_{fp}+\delta_{pv}.
\label {ne8}
\end{equation}
It is convenient to divide the numerator and denominator of the last line of (\ref{ne6}) by the  total number of atmospheric molecules $N_{\rm at}$ from (\ref{m16}) to find
\begin{eqnarray}
\delta_{nv} =\frac{(C-C_0)\delta_{fv}+C_0\delta_{pv}}{C}.
\label {ne8a}
\end{eqnarray}
The contemporary concentration $C$ of atmospheric CO$_2$ and the preindustrial concentration $C_0$ are 
\begin{equation}
C = \frac{N}{N_{\rm at}}\quad\hbox{and}\quad C_0 = \frac{P}{N_{\rm at}}.
\label {ne10}
\end{equation}
The no-exchange limit of (\ref{ne6}) or (\ref{ne8a}) is simply the average of $\delta^{13}C=\delta_{pv} = -6.5\,\perthou$, the $\delta^{13}C$ value  for the $P$ atoms of atmospheric  carbon from natural sources, and  $\delta^{13}C=\delta_{fv} = -28\,\perthou$, the $\delta^{13}C$ value  for the  $N-P$ atoms of atmospheric  carbon from fossil fuels.  For the year 2026, the concentrations needed to evaluate  (\ref{ne8a}) are $C_0 = 280$ ppm and $C = 430$ ppm

\begin{eqnarray}
\delta^{13}{\rm C}=\frac{(430-280)(-28)+280(-6.5)}{430}\,\perthou=-14.0\,\perthou.
\label {ne12}
\end{eqnarray}
This is much smaller than the observed value, $\delta^{13}{\rm C}=-8.7\,\perthou$, of (\ref{if14a}).
\subsubsection{Infinitely fast exchange}
From inspection of (\ref{rv17}) we see that for a fixed decay rate $\gamma_1$ of the concentration $C$ to the preindustrial value, $C_0= 280$ ppm,  but  for an  infinitely fast exchange rate,
\begin{equation}
\gamma_2\to \infty\quad\hbox{when}\quad \gamma_{12}\to \infty.
\label {fe2}
\end{equation}
Then we see from (\ref{su6}) that
\begin{eqnarray}
\Delta N_2&=&\int_{-\infty}^t dt'e^{-\gamma_2 (t-t')}F(t+t'-t)\nonumber\\
&=&\int_{-\infty}^t dt'e^{-\gamma_2 (t-t')}\bigg[F(t)
+\frac{(t'-t)}{1!}\frac{dF}{dt}(t)+\frac{(t'-t)^2}{2!}\frac{d^2F}{dt^2}(t)+\cdots\bigg]\nonumber\\
&=&\frac{1}{\gamma_2}F(t)
-\frac{1}{\gamma_2^2}\frac{dF}{dt}(t)+\frac{1}{\gamma_2^3}\frac{d^2F}{dt^2}(t)-\cdots\nonumber\\
&\to& 0\quad\hbox{as}\quad \gamma_2\to \infty.
\label {fe4}
\end{eqnarray}
Using (\ref{fe4}) with (\ref{su14}) we see that for the limit of infinitely fast isotope exchange
\begin{eqnarray}
\delta^{13}{\rm C}\to\delta_{pv}= -6.5\,\perthou \quad \hbox{as}\quad \gamma_{12}\to \infty.
\label {fe6}
\end{eqnarray}
This is the preindustrial value, no matter how large a fraction of fossil-fuel carbon is added to the atmosphere.

According to (\ref{fe6}), 
an infinitely fast exchange rate with $\gamma_2\to \infty$ gives no Suess effect at all. But from (\ref{ne12}) we see that a vanishing exchange rate, $\gamma_{12}=0$, with $\gamma_2\to \gamma_1$, gives much too large a Suess effect. The real exchange rate $\gamma_{12}$ of (\ref{rv17})  has a value between zero and infinity.
\subsection{Discretizaton}
For model calculations with readily available observational data,  it is useful to introduce a series of discrete times 
\begin{eqnarray}
[t_0, t_1,t_2,\ldots,t_n] = [1850, 1851,1853,\ldots,2024],\label {d2}
\end{eqnarray}
The numbers on the right side of (\ref{d2}) are the ends of calendar years in the industrial era. We assume that fossil fuel emissions, summarized by the data of references\,\cite{emissions, Fried}, are small enough to neglect before the year $t_0 = 1850$, that is,  we assume $F(t') = 0$ for $t'<t_0$.
Dividing both sides of (\ref{su6}) by the total number of atmospheric molecules $N_{\rm at}$ of (\ref{m16}) gives  the concentration increments at the end of the year $t_j$, with  $j\ge 1$,
\begin{eqnarray}
\Delta C_{qj}=\frac{\Delta N_q(t_j)}{N_{\rm at}}&=&\int_{-\infty}^{t_j} dt' \,e^{-\gamma_q (t_j-t')} \frac{F(t')}{N_{\rm at}}\nonumber\\
&=&\sum_{k=1}^j\int_{t_{k-1}}^{t_k}\, dt' e^{-\gamma_q (t_j-t')} \frac{F(t')}{N_{\rm at}}\nonumber\\
&\approx&\sum_{k=1}^j e^{-\gamma_q (t_j-t_k)}\int_{t_{k-1}}^{t_k}\,dt'\frac{F(t')}{N_{\rm at}},\label {d4}
\end{eqnarray}
or 
\begin{eqnarray}
\Delta C_{qj}=\sum_{k=1}^j e^{-\gamma_q (t_j-t_k)}E_k.\label {d6}
\end{eqnarray}
The concentration increments due to the release of fossil-fuel carbon atoms to the atmosphere in the year $t_k$  are 
\begin{eqnarray}
E_k=\int_{t_{k-1}}^{t_k}\,dt'\frac{F(t')}{N_{\rm at}}.\label {d8}
\end{eqnarray}
Dividing both sides of (\ref{su4}) by the total number of atmospheric molecules, $N_{\rm at}$ of (\ref{m16}), setting 
\begin{equation}
C_j=\frac{N(t_j)}{N_{\rm at}}, \label {d10}
\end{equation}
using the definition  (\ref{m12}) of $C_0$ and (\ref{d4}) of $\Delta C_{qj}$  we find for $j\ge 1$,
\begin{eqnarray}
C_j&=&\Delta C_{1j} +C_0\nonumber\\
&=&\sum_{k=1}^j e^{-\gamma_1 (t_j-t_k)}E_k +C_0, 
\label {d12}
\end{eqnarray}
an expression given in reference \,\cite{HE}.

In like manner, we can use (\ref{su14}) to write the isotope delta values for $j\ge 1$ as
\begin{eqnarray}
\delta^{13}{\rm C}_j &=&\frac{\Delta C_{2j}}{C_j} +\delta_{pv}\nonumber\\
&=&\frac{1}{C_j}\sum_{k=1}^j e^{-\gamma_2 (t_j-t_k)}E_k  +\delta_{pv}. 
\label {d14}
\end{eqnarray}
Here $C_j$ was given by (\ref{d12}) and $\delta_{pv} = -6.5\,\perthou$ was given by (\ref{if12}).  For the special case $j=0$, which labels the beginning year of the industrial era, both the atmospheric concentrations of CO$_2$ and the isotope delta value have their preindustrial values,
\begin{equation}
C_0=280\hbox{ ppm}\quad\hbox{and}\quad \delta^{13}{\rm C}_0 =\delta_{pv} = -6.5\,\perthou.\label {d16}
\end{equation}

The model curve for CO$_2$ concentration $C$ of Fig. \ref{Cdel} was plotted from (\ref{d12})  with $C_0=280$ ppm, in accordance with (\ref{m14}), and with $\gamma_1=1/\tau_1=1/60$ y, which gave  a good fit to observed values of $C$. A slightly different time constant,  $\tau_1=50$ y was found in reference\,\cite{HE}, where a value of $C_0 = 290$ ppm was used. The model curve for $\delta^{13}$C in Fig. \ref{Cdel} was plotted from (\ref{d14}) and with $ \gamma_2=1/\tau_2=1/8$ y. This gives  a good fit to observed values of $\delta^{13}$C.

Using $\tau_1=60$ y and $\tau_2 = 8$ y, we can use the exchange rate $\gamma_{12}$ of of (\ref{rv17}) to write the exchange time $\tau_{12}$ as
\begin{equation}
\tau_{12}=\frac{1}{\gamma_{12}} = \frac{\tau_1\tau_2} {\tau_1-\tau_2}= 9.2\hbox{ y}.\label {d18}
\end{equation}
\begin{figure}[h]
\begin{centering}
%\postscriptscale{stop.eps}{1}
\includegraphics[height=80mm,width=.8\columnwidth]{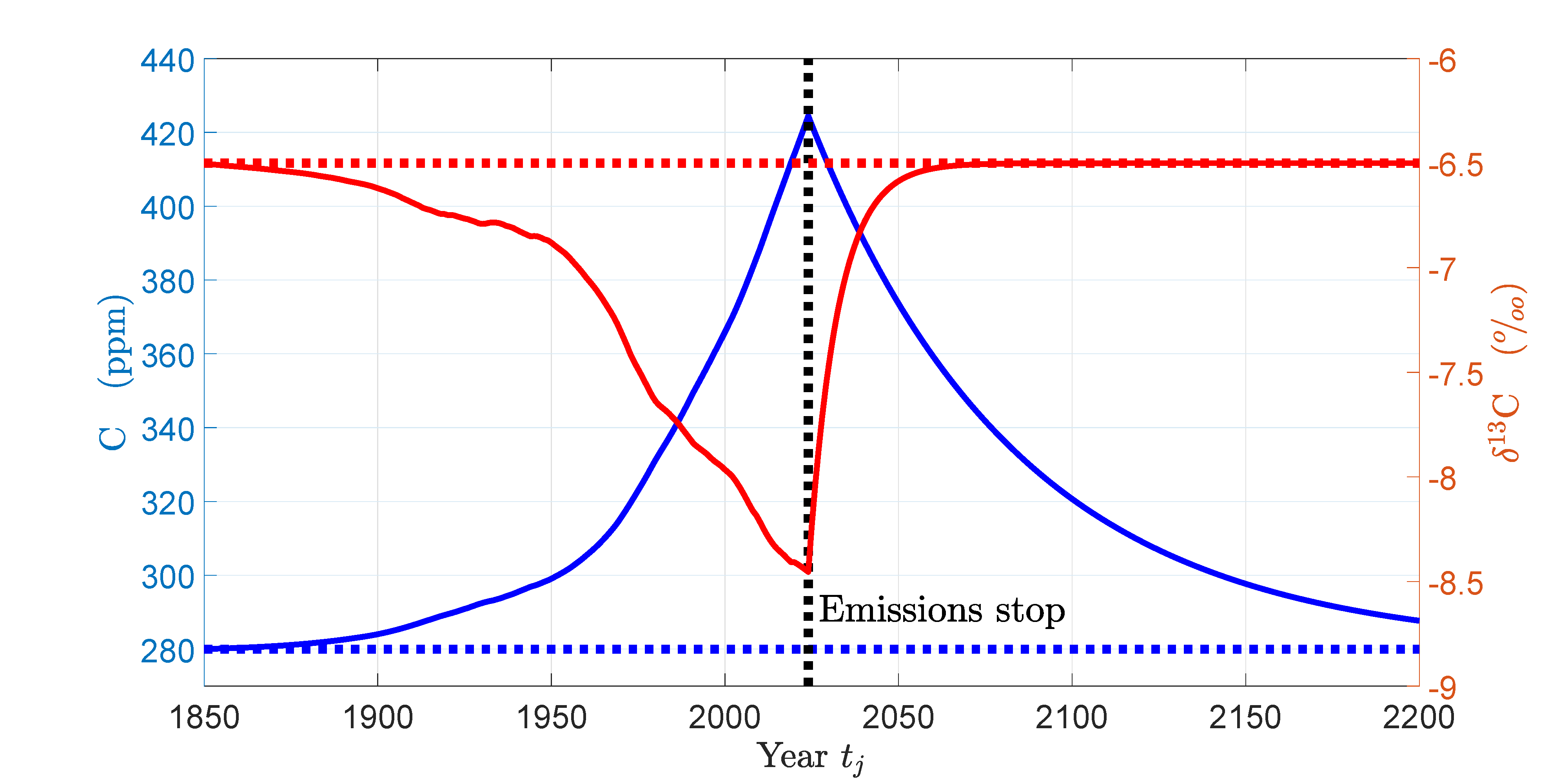}
\caption{Modelled time dependence of atmospheric CO$_2$ concentration $C$ and isotopic delta value $\delta^{13}$C if all fossil fuel emissions stopped after the year 2024, marked by the black dotted line.  Because of isotope exchange,  $\delta^{13}$C, shown as the smooth red line, evolves more quickly to its preindustrial value $\delta^{13}$C$_0 = -6.5 \,\perthou$, shown as the red dotted line, than does the atmospheric concentration $C$, shown as the smooth blue line, to its preindustrial value, $C_0 = 280$ ppm, shown as the blue dotted line.}
\label{stop}
\end{centering}
\end{figure}
\section{Complete halt of fossil fuel emissions}
To better understand the model, consider the hypothetical situation that all fossil fuel emissions stopped after the year 2024. Formally, we assume that the annual emissions (\ref{d8}) are
\begin{equation}
E_k=\left \{\begin{array}{l}\hbox{the values of Fig. \ref{emis} if $t_k\le 2024$, }\\
\hbox{$0$ if $t_k> 2024$.} \end{array}\right . 
\label{ch2}
\end{equation}
Substituting (\ref{ch2}) into (\ref{d12}) and (\ref{d14}) we find the modelled CO$_2$ concentrations $C$ and isotope delta values $\delta^{13}$C of Fig. \ref{stop}. 
Because of isotope exchange,  after fossil fuel emissions stop the isotope delta value $\delta^{13}$C relaxes more quickly to its preindustrial value $\delta^{13}$C$_0 = -6.5 \,\perthou$, than does the atmospheric concentration $C$ to its preindustrial value, $C_0 = 280$ ppm.
\section{Conclusions} The simple model discussed here gives good agreement with observations. It could be elaborated to include additional details, for example, the changes of $\delta^{13}{\rm C}$ from the beginning of the industrial era, when coal was the main fossil fuel to the present time when larger fractions of oil and natural gas are used \cite{Wang}. The fraction of radioactive $^{14}$C,  both that steadily produced by cosmic rays and the pulse introduced by above ground nuclear tests\, \cite{Graven} could be analyzed in a similar way.  

\appendix
\section{Appendix\label{a}}
\subsection{Some properties of $\delta^{13}C$}
In analogy to ({\ref{m4}), we define an isotope number vector for the VPDB (Vienna Pee Dee Belemnite) standard by
\begin{equation}
|V\rangle = \left [\begin{array}{l}V_1\\ V_2\end{array}\right]=V\left [\begin{array}{l}v_1\\ v_2\end{array}\right],\label {a2}
\end{equation}
where
\begin{equation}
 V=V_1+V_2,\quad v_1 = V_1/V,\quad v_2=V_2/V\quad \hbox{and}\quad v_1+v_2 = 1.\label {a4}
\end{equation}

For an arbitrary sample $X$ of carbon, with potentially different isotope ratios than that of the VPDB standard, the isotope number vector can be written as
\begin{equation}
|X\rangle = \left [\begin{array}{l}X_1\\ X_2\end{array}\right]=X\left [\begin{array}{l}x_1\\ x_2\end{array}\right],\label {a6}
\end{equation}
where
\begin{equation}
 X=X_1+X_2,\quad x_1 = X_1/X,\quad x_2=X_2/X\quad \hbox{and}\quad x_1+x_2 = 1.\label {a8}
\end{equation}

Then we can write the parameter $\delta^{13}C$ of (\ref{if10})  as
\begin{eqnarray}	
\delta_{xv}&=& \frac{x_2/x_1}{v_2/v_1} -1 \nonumber\\
&=& \frac{x_2(1-v_2)}{v_2(1-x_2)}-1.
\label{a8}
\end{eqnarray}
Solving the second line of (\ref{a8}) for $x_2$ in terms of $v_2$ and $\delta_{xv}$, and setting $x_1=1-x_2$  we find
\begin{equation}	
x_2=v_2+\frac{\delta_{xv}v_1v_2}{1+\delta_{xv}v_2}\quad\hbox{and}\quad 
x_1=v_1-\frac{\delta_{xv}v_1v_2}{1+\delta_{xv}v_2}
\label{a10}
\end{equation}
From (\ref{a8}) we see that the isotope delta values of a carbon samples, $X$ compared to itself is zero,
\begin{equation}	
\delta_{xx}=0.
\label{a12}
\end{equation}
From (\ref{a8}) we find the {\it reciprocal law} for isotope delta values for two carbon samples, $X$ and $Y$,
\begin{equation}	
(1+\delta_{xy})(1+\delta_{yx})=1.
\label{a14}
\end{equation}
From (\ref{a8}) we also find the {\it addition law} for isotope delta values for three carbon samples, $X$, $Y$ and $V$, 
\begin{equation}	
(1+\delta_{yx})(1+\delta_{xv})=1+\delta_{yv}.
\label{a16}
\end{equation}
or 
\begin{eqnarray}	
\delta_{yv} &=& \delta_{yx}+\delta_{xv} + \delta_{yx}\delta_{xv} \nonumber\\
 &\approx& \delta_{yx}+\delta_{xv}.
\label{a18}
\end{eqnarray}
Since the isotope delta values of natural samples of carbon are on the order of  a per cent or less, they are approximately additive, as shown by the second line of (\ref{a18}).
\section*{Acknowledgements}
The authors would like to thank their respective universities, York and Princeton, for support.


\begin{thebibliography}{99} 
\bibitem{EHB}F. Engelbeen, R. Hannon and D. Burton, {\it The Human Contribution to Atmospheric Carbon Dioxide}, (2024).
\url{https://co2coalition.org/wp-content/uploads/2024/12/Human-Contribution-to-Atmospheric-CO2-digital-compressed.pdf}

\bibitem{emissions} Global Carbon Emissions,\\  \url{https://globalcarbonbudget.org/download/2371/?tmstv=176281725}

\bibitem{Fried} P. Friedlingstein et al, {\it Global Carbon Budget 2025}, Earth System Sci. Data, {\bf 18}, 3211-3288 (2026).

\bibitem{Lan} X. Lan and R. Keeling, {\it Trends in CO$_2$, CH$_4$, N$_2$O, SF$_6$: Trends in Atmospheric Carbon Dioxide: Mauna Loa CO$_2$ Annual Mean Data}. NOAA Global Monitoring Laboratory, Boulder, CO, USA, (2026). \url{https://gml.noaa.gov/webdata/ccgg/trends/co2/co2_annmean_mlo.txt}

\bibitem{Etheridge}D. M. Etheridge et al, {\it Natural and Anthropogenic Changes in Atmospheric CO$_2$ Over the Last 1000 Years from Air in Antarctic Ice and Firn}, J. Geophys. Res. {\bf 101} (D2), 4115–4128 (1996).

\bibitem{Keeling0} R. F. Keeling, S. C. Piper, A. F. Bollenbacher and S. J. Walker, {\it Ice-Core Merged Products: Yearly Spline. Scripps CO$_2$ Program}, Scripps Institution of Oceanography, University of California, La Jolla, CA, USA. \url{https://keelinglabsites.ucsd.edu/websitedataco2/spline_merged_ice_core_yearly.csv}

\bibitem{MacFarling} C. MacFarling Meure et al, {\it Law Dome CO$_2$, CH$_4$ and N$_2$O Ice Core Records Extended to 2000 Years BP}, Geophys. Res. Lett. {\bf 33}, L14810 (2006).

\bibitem{Rubino1} M. Rubino et al, {\it Law Dome, Antarctica 2000 Year Ice Core CO$_2$, CH$_4$, N$_2$O and $\delta^{13}$C-CO$_2$ Data}, World Data Service for Paleoclimatology, Boulder, CO, USA, and NOAA Paleoclimatology Program, National Centers for Environmental Information (NCEI), (2024).
 \url{https://www.ncei.noaa.gov/pub/data/paleo/icecore/antarctica/law/law2018d13c-co2-noaa.txt}

\bibitem{Rubino2} M. Rubino et al, 
{\it Revised Records of Atmospheric Trace Gases CO$_2$, CH$_4$, N$_2$O, and $\delta^{13}$C CO$_2$ over the last 2000 years from Law Dome, Antarctica},
Earth System Science Data, {\bf 11} (3), 473–492 (2019). \url{https://doi.org/10.5194/essd-11-473-2019}

\bibitem{Keeling} R. F. Keeling, E. J. Morgan and C. D. Keeling, {\it Atmospheric Flask CO$_2$ and Isotopic Data Sets – Multiple Sampling Stations (Archive 2024-04-19)},  Scripps CO$_2$ Program Data, UC San Diego Library Digital Collections. \url{ https://doi.org/10.6075/J0HQ3X30}

\bibitem{Alexander} D. Alexander, J. D. Ferguson, A. Glatzle, W. Happer and W. A. van Wijngaarden, {\it Livestock, Methane and Climate}, Atmos. and Oceanic Phys. arXiv: 2601.18522v1 (2026). 
\url{https://co2coalition.org/publications/livestock-methane-and-climate/}

\bibitem{HE} W. Happer, G. Wrightstone, and F. B. Soepyan, {\it Human Emissions and Atmospheric Concentrations of Carbon Dioxide}, CO$_2$ Coalition, (2026). 
\url{https://co2coalition.org/wp-content/uploads/2026/08/Emission-Concentration-2026-08-21-1.pdf}

\bibitem{Newton} Newton's law of cooling.\\
\url{https://www.quadco.engineering/en/know-how/newtons-law-of-cooling.htm}

\bibitem{IF} Isotope fractionation. \url{https://en.wikipedia.org/wiki/Isotope_fractionation}

\bibitem{PD} Pee Dee Belemnite. \url{https://pubmed.ncbi.nlm.nih.gov/35312289/}

\bibitem{aPD}P. J. H. Dunn et al, {\it Redetermination of R(\,$^{13}$C/\,$^{12}$C) for Vienna Peedee Belemnite}, Rapid Comm. in Mass Spectrometry, John Wiley, (2024). 
\url{https://analyticalsciencejournals.onlinelibrary.wiley.com/doi/10.1002/rcm.9773}

\bibitem{Suess} The Suess Effect.\\
\url{https://en.wikipedia.org/wiki/Suess_effect}

\bibitem{Wang} P. Wang et al, {\it Stable carbon isotopic characteristics of fossil fuels in China}, Science of Total Environ., {\bf 805} (2022). \url{https://doi.org/10.1016/j.scitotenv.2021.150240}

\bibitem{Graven} H. Graven, R. E. Keeling and J. Rogelj, {\it Changes to Carbon Isotopes in Atmopspheric CO$_2$ Over the Industrial Era and Into the Future}, Global Biochemical Cycles, {\bf 34}, (2019). \\
\url{https://doi.org/10.1029/2019GB006170}.

\end{thebibliography}
\end{document}